\documentclass{article}
\usepackage[utf8]{inputenc}
\usepackage{authblk}
\usepackage[numbers,sort&compress]{natbib}
\usepackage{geometry}
\usepackage{multicol}
\usepackage{hyperref}
\usepackage{gensymb}
\usepackage{float}
\UseRawInputEncoding
\usepackage{tabularx}
\usepackage{graphicx}
\usepackage{caption}
\usepackage{subcaption}
\usepackage{comment}
\usepackage[symbol]{footmisc}
\usepackage{xcolor}
\begin{document}

\title{Structural Anisotropy in Sb Thin Films}
\author[1]{Pradip Adhikari}
\author[1]{Anuradha Wijesinghe}
\author[1]{Anjali Rathore}
\author[2]{Timothy Jinsoo Yoo}
\author[1,3]{Gyehyeon Kim}
\author[1,3]{Hyoungtaek Lee}
\author[4]{Sinchul Yeom}
\author[4]{Alessandro R. Mazza}
\author[3]{Changhee Sohn}
\author[3]{Hyeong-Ryeol Park}
\author[4]{Mina Yoon}
\author[4]{Matthew Brahlek}
\author[2]{Honggyu Kim}
\author[1,*]{Joon Sue Lee}
\setcounter{Maxaffil}{0}
\renewcommand\Affilfont{\itshape\small}
\affil[1]{Department of Physics and Astronomy, University of Tennessee, Knoxville, Tennessee 37996, USA}
\affil[2]{Materials Science and Engineering, Herbert Wertheim College of Engineering, University of Florida, Gainesville, Florida 32611, USA }
\affil[3]{Department of Physics, Ulsan National Institute of Science and Technology, Ulsan, 44919 South Korea}
\affil[4]{Materials Science and Technology Division, Oak Ridge National Laboratory, Oak Ridge, Tennessee 37831, USA}
\affil[*]{E-mail: jslee@utk.edu}
\date{}

\maketitle

\begin{abstract}
Sb thin films have attracted wide interests due to their tunable band structure, topological phases, and remarkable electronic properties. We successfully grow epitaxial Sb thin films on a closely lattice-matched GaSb(001) surface by molecular beam epitaxy. We find a novel anisotropic directional dependence of their structural, morphological, and electronic properties. The origin of the anisotropic features is elucidated using first-principles density functional theory (DFT) calculations. The growth regime of crystalline and amorphous Sb thin films was determined by mapping the surface reconstruction phase diagram of  the GaSb(001) surface under Sb$_2$ flux, with confirmation of structural characterizations. Crystalline Sb thin films show a rhombohedral crystal structure along the rhombohedral (104) surface orientation parallel to the cubic (001) surface orientation of the GaSb substrate. At this coherent interface, Sb atoms are aligned with the GaSb lattice along the [$\bar1$10] crystallographic direction but are not aligned well along the [110] crystallographic direction, which results in anisotropic features in reflection high-energy electron diffraction patterns, surface morphology, and transport properties. 
Our DFT calculations show that the anisotropic features originate from the GaSb surface, where Sb atoms align with the Ga and Sb atoms on the reconstructed surface. The formation energy calculations confirm that the stability of the experimentally observed structures. Our results provide optimal film growth conditions for further studies of novel properties of Bi$_{1-x}$Sb$_x$ thin films with similar lattice parameters and an identical crystal structure as well as functional heterostructures of them with III-V semiconductor layers along the (001) surface orientation, supported by a theoretical understanding of the anisotropic film orientation. 
\end{abstract}

\section{Introduction}
Group-VA elemental thin films (phosphorus, arsenic, antimony, and bismuth) have gained significant attention in recent years due to rich and promising properties such as high carrier mobilities, outstanding optical and thermodynamic responses, tunable band gap, and non-trivial topological phases \cite{Zhang2018RecentExperiment, Wu2020ElectricalLayers, Gui2019Two-dimensionalApplications, Chen2020REVIEW2020}. Among the group-VA elements, Sb and Bi are relatively heavy elements with a strong spin-orbit coupling. Multiple topological phases in Sb and Bi thin films, including quantum spin Hall insulator phase in the two-dimensional (2D) limit, three-dimensional (3D) topological insulator (TI) phase, and 3D higher-order TI phase, have been theoretically proposed, and some of the features have been experimentally demonstrated\cite{Zhang2012TopologicalEffectb, Lima2015TopologicalBilayers, Murakami2006QuantumCoupling,Wang2015TopologicalAdsorption}. In general, electronic band structures with non-trivial topology can be modified by strain, electric and magnetic fields, and thickness. In Sb thin films, it is theoretically predicted that the quantum confinement effect opens up a bulk band gap when film thickness is less than 7.8 nm where it enters into the 3D TI regime. Going even below a certain thickness transforms the topological phase into the quantum spin Hall state because of the surface coupling effect \cite{Zhang2012TopologicalEffectb}. Antimonene, Sb analog of graphene, is a 2D hexagonal lattice of Sb atoms. In addition to the non-trivial topology of antimonene (quantum spin Hall state), remarkable properties including stability in air, high electron mobility, and thermoelectric and ferroelectric properties have attracted wide interests \cite{Ares2017RecentMaterial, Wang2019Antimonene:Application, Liu2020Antimonene:Material,Xue2021RecentGrowth}. 

Precise control over the film thickness is critical to investigate the quantum confinement effect in Sb thin films, and molecular beam epitaxy (MBE) is advantageous for layer-by-layer construction of topological quantum materials \cite{Brahlek2020TopologicalEpitaxy}. By using MBE, Sb thin films have been synthesized on various substrates since 1980s. Early studies of Sb films grown on GaAs(110), InP(110), and InP(001) focused on the use of Sb as a capping layer or a Schottky barrier \cite{Dumas1992Sb-cappingGaSb100, Clark1995AntimonyInAlSb100,Cheng1984Schottky-barrierArsenide, Kwok1987SchottkySilicon}. Moreover, deposition of Sb on direct band-gap semiconductors of InSb(111) and GaSb(111) has been investigated for the purpose of development of superlattices with an indirect narrow gap/direct gap heterostructures \cite{Gaspe2013EpitaxialWells}. Recent reports on epitaxial Sb mostly focus on demonstration of ultrathin Sb films or antimonene layers for their novel 2D nature and topologically non-trivial properties. Due to the hexagonal lattice structure of antimonene, van der Waals 2D substrates such as graphene \cite{Sun2018VanGraphene}, as well as (111) surface orientation of copper \cite{Niu2019ModulatingDesign} have been used for MBE growth of Sb layers. The most common crystalline structure for group V elemental solids is the rhombohedral structure, thus also for Sb. Under ultrahigh vacuum (UHV) environment, rhombohedral Sb(111) layers can be epitaxially grown on closely lattice-matched GaSb(111) with hexagonal lattice arrangement  \cite{Dura1995EpitaxialHeteroepitaxy, Gaspe2013EpitaxialWells}. However, on cubic (001) surface orientation, polycrystalline nucleation of Sb with rough surfaces was reported with no success in epitaxial growth of Sb layers  \cite{Dura1995EpitaxialHeteroepitaxy}.

In this work, we report wide-area Sb thin films coherently grown on cubic GaSb(001) surface by MBE. We carefully study surface kinetics and crystalline phase of Sb on closely lattice-matched cubic GaSb(001) surface, under UHV environment. We first delve into the surface kinetics of the GaSb(001) surface in the presence of Sb flux over a wide range of temperature from 450$\degree$C down to room temperature and find nucleation conditions of Sb films. We employ \emph{in-situ} reflection high-energy electron diffraction (RHEED) patterns to observe surface reconstruction on the GaSb(001) surface, as well as abrupt changes occurring at the surface when Sb layers starts to grow. We successfully grew crystalline Sb thin films coherent to the GaSb(001) atomic structures below 120$\degree$C. 
The Sb structure turned out to be rhombohedral along the (104) surface orientation parallel to the cubic (001) surface orientation of the GaSb substrate, confirmed by x-ray diffraction (XRD) and electron diffraction using transmission electron microscopy (TEM). Our DFT calculations show that the anisotropic features in Sb thin films, which refer to the directional dependence of their structural and morphological properties, originate from the reconstruction of the GaSb surface.No cubic phase of Sb was seen from any of the grown films, consistent with the unstable cubic phase of Sb at ambient conditions \cite{Donohue1974TheElements.}. The observed Sb(104) planes are aligned to the GaSb(001) lattices along the [$\bar1$10] direction, whereas mismatched lattices are expected along the [110] direction. This anisotropic lattice matching of the rhombohedral Sb(104) and cubic GaSb(001) layers result in 1) spottiness of RHEED along the [110] crystalline direction, 2) elongated formation of Sb structures along the [$\bar1$10] crystalline direction, observed by scanning electron microscopy (SEM) and atomic force microscopy (AFM), and 3) anisotropic transport with relatively lower resistance along the [$\bar1$10] crystalline direction in comparison to the [110] direction. The observed anisotropic features can be significantly reduced by growing Sb films at lower temperatures. Our DFT calculations show that the Sb(104) layers with observed anisotropy are stable due to the (1 $\times$ 3) surface reconstruction of GaSb(001) surface. The successful demonstration of coherent, rhombohedral Sb thin films grown on cubic GaSb(001) substrates paves the way to embed crystalline Sb layers into well-developed and widely-used cubic semiconductor substrates for fundamental studies of topological nature of Sb thin films as well as for applications using the remarkable electronic, optical, and thermoelectric properties. This study can be further extended to studies of Bi$_{1-x}$Sb$_x$ thin films on lattice-matched cubic substrates. Bi$_{1-x}$Sb$_x$ has shown multiple topological phases, which have potential applications in spintronics and quantum computing \cite{Hasan2010Colloquium:Insulators, HuynhASwitching, Mellnik2014Spin-transferInsulator}.

\section{Results and Discussion}
\subsection{Surface reconstruction and Sb film growth on GaSb(001) surface}
To achieve optimal growth conditions of Sb thin films, surface reconstruction phase diagram of the GaSb(001) surface was investigated. In ultrahigh vacuum chamber, the native oxide on the GaSb(001) surface was thermally desorbed, confirmed by the appearance of RHEED patterns, in the presence of Sb$_2$ flux. GaSb desorption temperature of 540$\degree$C was used to calibrate the pyrometer. On the desorbed surface, a GaSb homoepitaxial buffer layer was grown at 450$\degree$C and streaky $(1 \times 3)$ RHEED patterns confirmed the smooth surface under Sb-rich condition.  To obtain the surface reconstruction phase diagram of GaSb in the presence of Sb$_2$ flux, the Sb$_2$ flux was kept constant, and change in the RHEED patterns was tracked with decrement of substrate temperature. When the RHEED patterns significantly changed with deposition of Sb layer at lower temperatures, substrate temperature was raised above 400$\degree$C until GaSb (1 $\times$ 3) RHEED patterns reappeared, and a thin GaSb layer was grown to obtain a smooth surface. This process was repeated with a change in the Sb$_2$ flux.

\begin{figure}[t!]
    \centering
    \includegraphics[width= \textwidth, height = 13 cm]{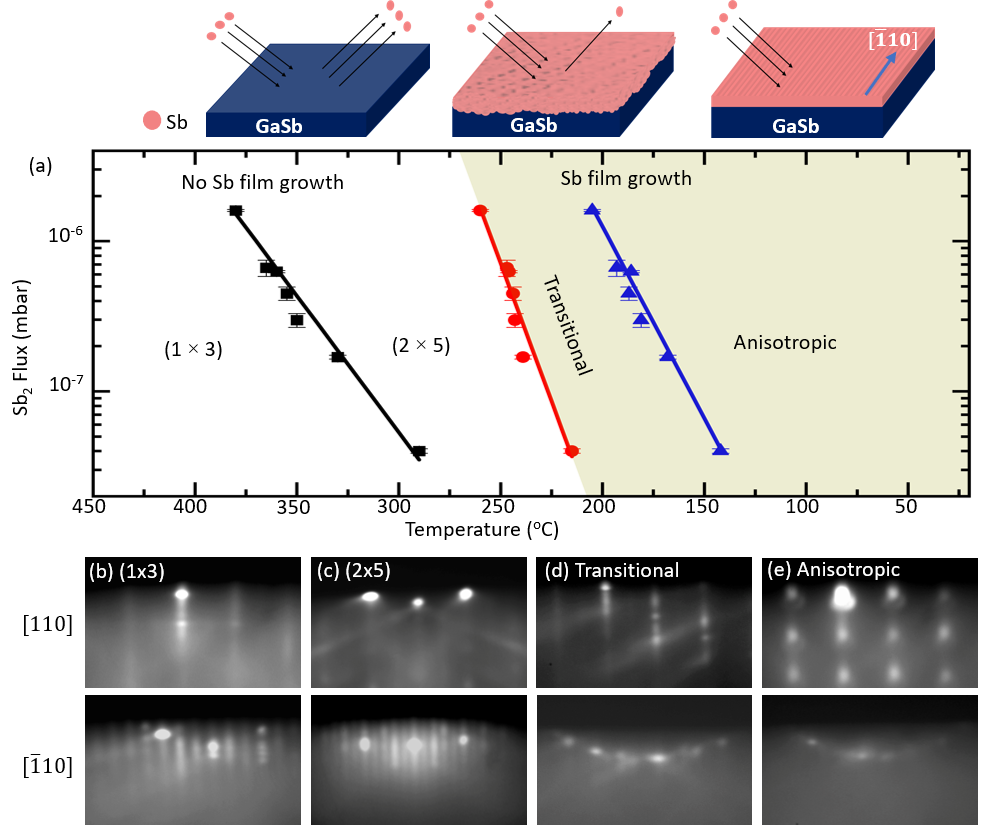}
    \caption{(a) Surface reconstruction phase diagram for GaSb(001) in the presence of Sb$_2$ flux and Sb film growth illustration showing no growth, amorphous Sb with rough interface, and anisotropic crystalline Sb film with clean interface on GaSb(001) as growth temperature decreases.  
    Representative RHEED images observed in (b) (1 $\times$ 3), (c) (2 $\times$ 5), (d) transitional, and (e) anisotropic RHEED regions. The shaded part indicates the region of Sb film growth. Error bars is the standard error of mean from flux measurement on beam flux monitor. The upper and lower rows show the RHEED images in the [110] direction and the [$\bar1$10] direction, respectively. }
    \label{phase}
\end{figure}

Figure \ref{phase}(a) shows the phase diagram for the surface reconstruction of the GaSb(001) surface under Sb$_2$ flux in the substrate temperature range from 450$\degree$C down to room temperature. The flux values are expressed in units of beam equivalent pressure (mbar) as measured by beam flux monitor of a Bayard-Alpert ionization gauge. The temperature values above 270$\degree$C were measured using a pyrometer focused on the sample while lower temperatures were from a thermocouple attached to a manipulator holding the sample on a tungsten sample holder. The data points on the plots indicate transitions of the RHEED patterns. The [(1 $\times$ 3) $\rightarrow$ (2 $\times$ 5)] transition is the GaSb(001) surface reconstruction, consistent with previous reports \cite{Bracker2000SurfaceGaSb,Nouaoura1997ModificationFlux}, indicating there is no Sb film grown on the surface. The RHEED pattern became blurry and dimmer between 200 $\degree$C and 250 $\degree$C, depending on the Sb$_2$ flux, indicating Sb atoms start to stick to the surface. Upon decreasing the substrate temperature, RHEED exhibited a sudden alteration, displaying distinctively spot-like patterns in the [110] direction and relatively indistinct but still streaky patterns in the [$\bar1$10] direction. This implies that in the [110] direction, the electron beam detected three-dimensional nanostructures, while in the [$\bar1$10] direction, it did not detect any significant three-dimensional features. This anisotropic spotty/streaky RHEED feature can be obtained with elongated 3D nanostructures on the surface, which is the case of the Sb thin films on the GaSb(001) surface as confirmed by surface morphology characterizations. Sb thin films grown in the transitional and the anisotropic RHEED regions were further characterized in the following sections.

\subsection{Crystal structure of Sb thin films}
\begin{figure}[t!]
    \centering
    \includegraphics[width=0.5\textwidth]{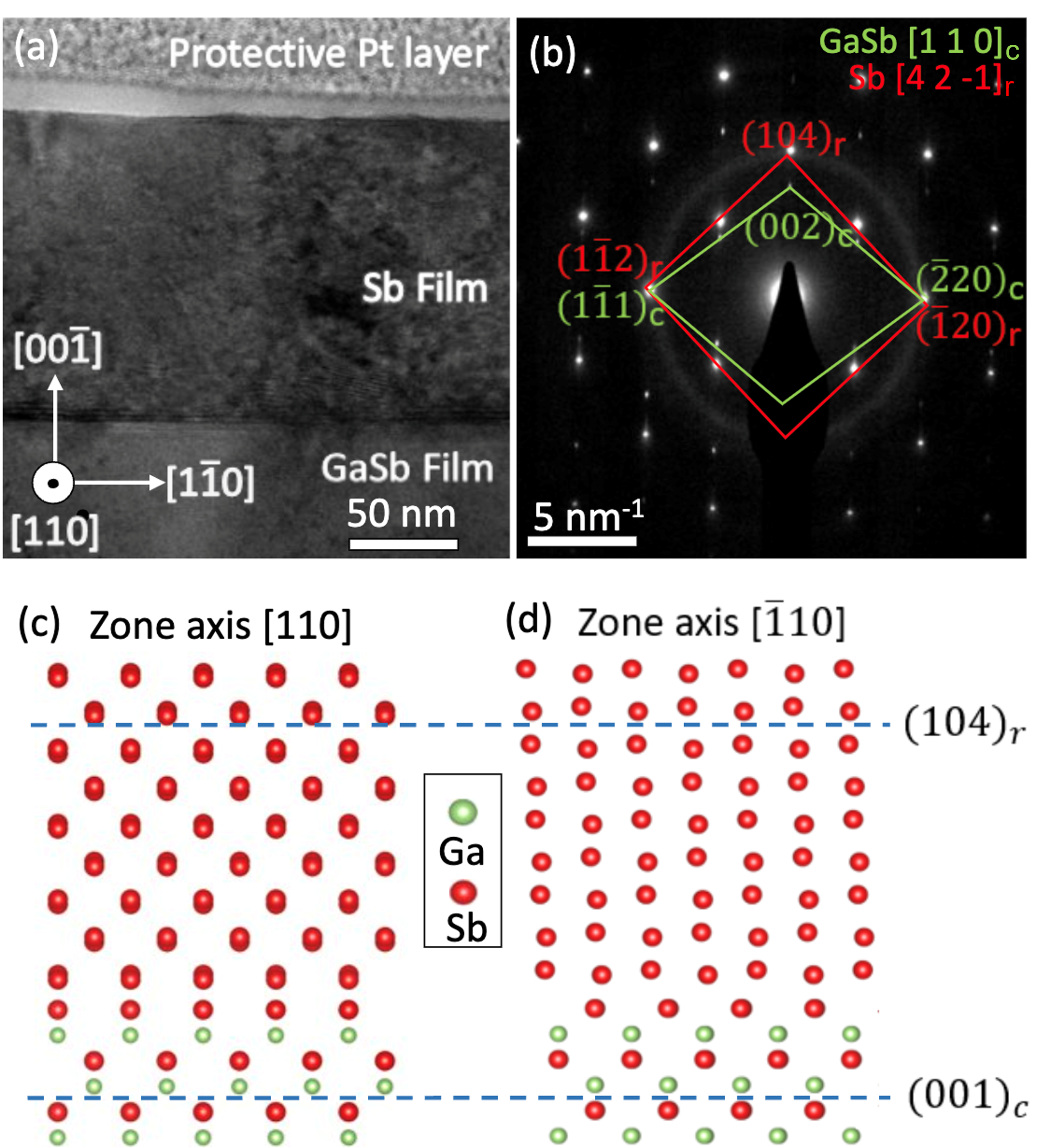}
    \caption{(a) HRTEM image of film stack for Sb film grown at 120\degree C. (b) SAED pattern acquired at the interface between GaSb buffer layer and Sb film. The diffraction spots for each structure are outlined in the red and green diamonds for Sb and GaSb, respectively. The (001)$_c$ growth plane of GaSb aligns with the (104)$_r$ growth plane of the Sb film. Illustrated crystal structure of rhombohedral Sb thin film on cubic GaSb layer, as seen in (c) along the [110]$_c$ zone axis and (d) along the [$\bar1$10]$_c$ zone axis.}
    \label{C_TEM}
\end{figure}

To investigate the crystal structure of Sb thin films,  two Sb films were prepared in the transitional and anisotropic RHEED regions based on the surface reconstruction phase diagram study. Two samples were grown at a manipulator temperature of 120$\degree$C in the anisotropic RHEED region and at 250 $\degree$C in the transitional RHEED region, respectively, with identical Sb$_2$ flux of 7.38 $\times 10  ^{-7}$ mbar and growth time. While cooling down the substrate after GaSb buffer layer growth at 500$\degree$C, Sb$_2$ flux was closed below 400 $\degree$C to prevent any unattended Sb growth and reopened after the substrate reached the desired temperatures. 
 High-resolution TEM (HRTEM) on a cross-section of the sample grown at 120$\degree$C shows the GaSb buffer layer, Sb film, and the protective Pt layer deposited by focused ion beam \{Fig. \ref{C_TEM}(a)\}. 
  
 The interface between Sb and GaSb is abrupt. Through selected area electron diffraction (SAED) analysis \{Fig. \ref{C_TEM}(b)\}, the Sb film is determined to adopt the rhombohedral phase with a growth plane of (104)$_r$, where the subscript indicates the rhombohedral structure. The Sb growth plane of (104)$_r$ is parallel to GaSb growth plane of (001)$_c$ of the cubic structure, which is later confirmed with the XRD measurement as well. Figures \ref{C_TEM}(c) and (d) illustrate the proposed growth orientation of the rhombohedral Sb on cubic GaSb according to these findings. Along the [$\bar1$10] direction of GaSb, Sb atoms in the Sb film align well with Ga and Sb atoms in the GaSb layer, whereas the positions of the atoms in the two layers do not match along the [110] direction. 

\begin{figure}[t!]
    \centering
    \includegraphics[width= 0.65\textwidth]{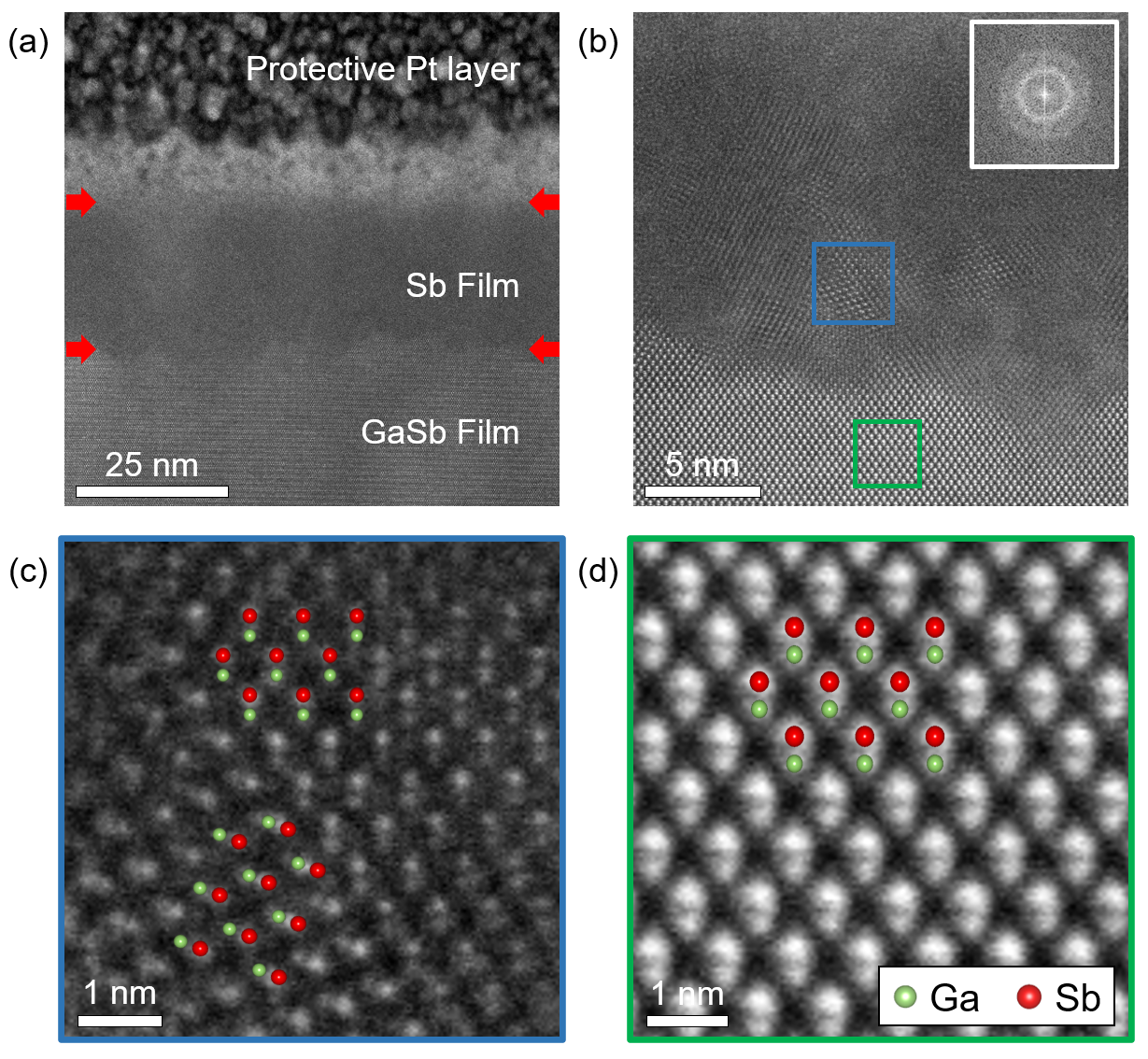}
    \caption{(a) HAADF-STEM image of film stack for Sb film grown at 250$\degree$C. (b) High-magnification HAADF-STEM image near the interface between the Sb film and GaSb buffer layer, showing that the interface is rough. Top right inset is a fast Fourier transform (FFT) image of the Sb film away from the interface showing ring patterns, indicative of an amorphous structure. (c) Magnified image of one of several crystalline GaSb patches observed in the Sb film from the blue box in (b). (d) Magnified image of the GaSb buffer layer viewed along [110] from the green box in (b). }
    \label{A_TEM}
\end{figure}

In contrast to the rhombohedral Sb thin film in the anisotropic RHEED region, the sample grown at 250$\degree$C, in the transitional RHEED region turns out to be amorphous, and the interface between Sb and GaSb is rougher than that of sample grown at 120 $\degree$C \{see Fig. \ref{A_TEM}\}.  A high-magnification high-angle annular dark-field scanning TEM (HAADF-STEM) image shows some crystallinity in the Sb film up to about 5 nm above the interface \{Fig. \ref{A_TEM}(b)\}, but it is mainly amorphous beyond that region. It is likely that in the transitional RHEED region, Ga atoms from GaSb layer diffuse into the Sb film, forming GaSb patches within the Sb film, as seen in Fig. \ref{A_TEM}(c).The orientations of the crystalline GaSb patches are different from each other, which is distinctive from the well-oriented single-crystalline GaSb layer below the interface. By using the same growth conditions (Sb flux and growth time), the thickness of sample grown at 250$\degree$C is around 25 nm whereas sample grown at 120$\degree$C is much thicker (100 nm). This indicates that in the transitional RHEED region, Sb atoms are partially desorbed and partially deposited on the GaSb surface.  No Sb peak was observed on the XRD of sample grown at 120$\degree$C, consistent with the amorphous nature observed by HAADF-STEM imaging. 

\begin{figure}[t!]
    \centering
    \includegraphics[width= \textwidth]{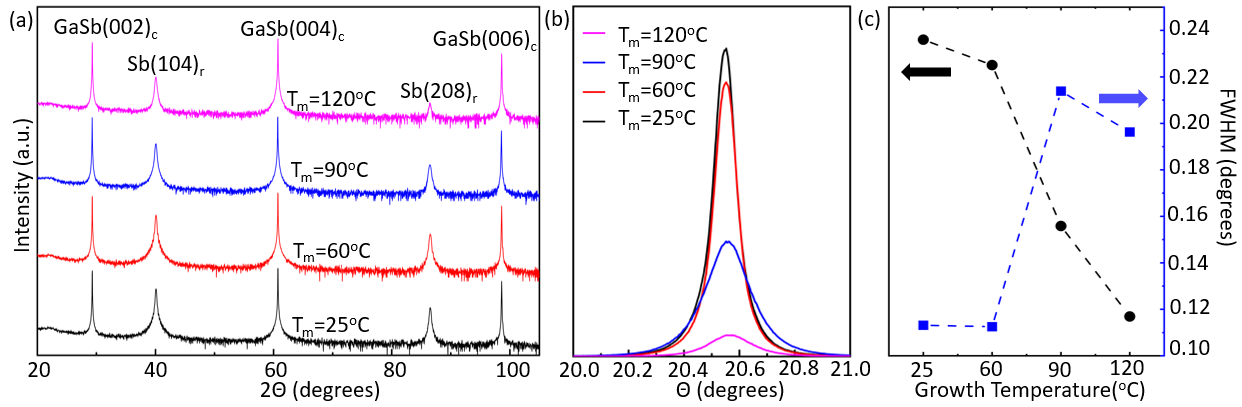}
    \caption{(a) XRD for samples grown at 25$\degree$C, 60$\degree$C, 90$\degree$C, and 120$\degree$C. The peaks appearing at 2$\theta$ values of 40$\degree$ and 82$\degree$ correspond to the Sb (104)$_r$ and (208)$_c$ planes while the other peaks are emerging from the GaSb(001) substrate. (b) Rocking curves for the samples grown at 25$\degree$C, 60$\degree$C, 90$\degree$C, and 120$\degree$C at  2$\theta$ value of 41$\degree$. The peaks for samples grown at lower temperatures are taller and sharper, while samples grown at higher temperatures give broader and shorter peaks, as plotted in (c) Sb peak intensity (black circles) and FWHM (blue squares) of the samples grown at four different temperatures. Dashed lines are guides to the eye.}
    \label{xrd}
\end{figure}

To further achieve the optimal quality of the rhombohedral Sb thin films, four different samples expecting equal thickness of approximately 50 nm were grown with constant Sb flux of 7.38 $\times 10  ^{-7}$ mbar and manipulator temperatures (T$_m$) of 25\degree C, 60\degree C, 90\degree C and 120\degree C, respectively. The $(1 \times 3)$ surface reconstruction of GaSb was seen before the growth of the Sb films. Figure \ref{xrd}(a) shows XRD peaks of the samples and all the films showed the crystalline nature of Sb. The peaks found at two values around 41$\degree$ and 82$\degree$ correspond with the Sb(104)$_r$ and Sb(208)$_r$ planes, respectively. The three tall peaks present in all the plots at $2 \theta$ angles of 29$\degree$, 61$\degree$ and 98$\degree$ correspond with the peaks of GaSb(002)$_c$, (004)$_c$ and (006)$_c$. The XRD results confirm that Sb grows in the rhombohedral structure with (104)$_r$ plane, which is parallel to the GaSb(001)$_c$ surface. The widths and heights of Sb XRD peaks vary depending on the growth temperatures. Particularly, the height of Sb(208)$_r$ peak decreases as the growth temperature increases. A rocking curve scan for each sample at $2 \theta$ = 41$\degree$ with identical measurement conditions revealed Gaussian-shaped rocking curve for all the samples. Consistent with the Sb XRD peak height variation, out of the four rocking curves, peaks from samples grown at 25$\degree$C and 60$\degree$C are significantly taller than those of samples grown at 90$\degree$C and 120$\degree$C, and Sb film grown at highest temperature shows the shortest peak \{Figs. \ref{xrd}(b) and (c)\}. Full-width half maxima (FWHM)) in Sb films grown at lower temperatures are narrower in comparison to those grown at higher temperatures, indicating higher film quality with fewer defects and curvature in films, consistent with the electron microscopy and RHEED results. 

\subsection{Anisotropic surface morphology}
\begin{figure}[t!]
    \centering
    \includegraphics[width=\textwidth]{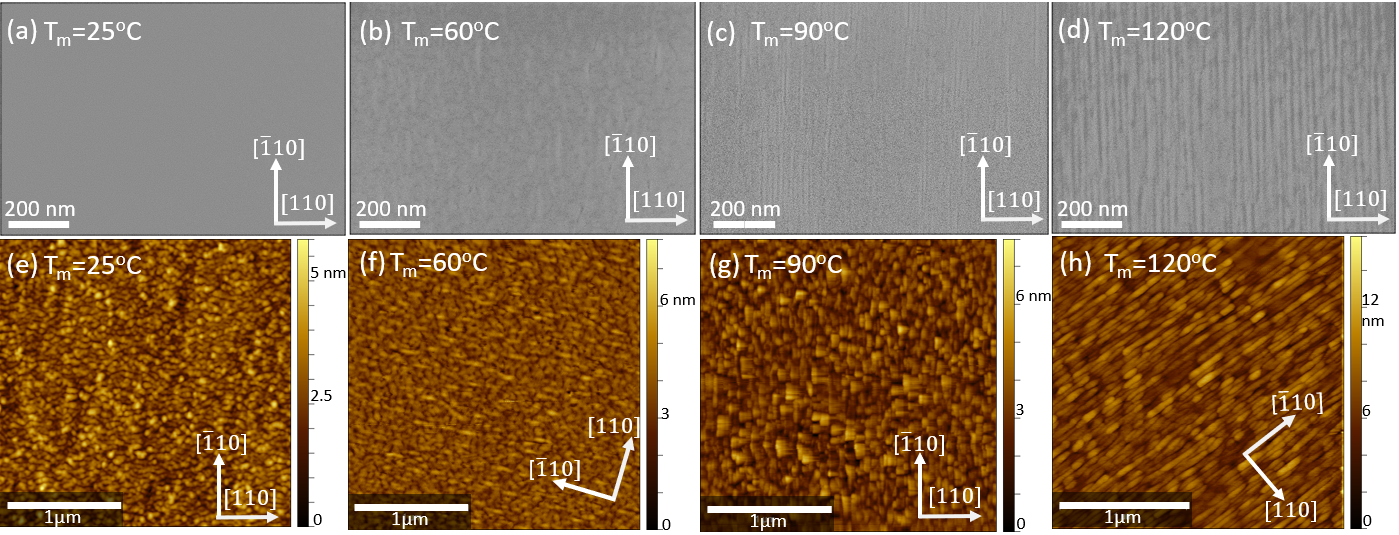}
    \caption{(a-d) SEM and (e-h) AFM images for the samples grown at 25$\degree$C, 60$\degree$C, 90$\degree$C, and 120$\degree$C. Roughness of the grown film increases with the growth temperature. Line-like features (elongated Sb structures) along [$\bar{1}10$] direction is more distinct on the samples grown at higher temperatures.}
    \label{semafm}
\end{figure}

Surfaces of the samples grown at four different temperatures were scanned using SEM and AFM. SEM images in {Figs. \ref{semafm}(a) through (d)} show line-like features along the [$\bar{1}$10] direction in the samples grown at 60$\degree$C, 90$\degree$C, and 120$\degree$C while no prominent features were seen in sample grown at 25$\degree$C. The contrast of these lines appear to get stronger with increasing growth temperatures, which are consistent with the AFM results. The sample grown at 25$\degree$C shows the smoothest surface with no clear tendency of directional structures whereas slight elongation of nanostructures on the top surface starts to be seen from the sample grown at 60$\degree$C \{Figs. \ref{semafm}(e) and (f)\}. Elongated Sb structures are more prominent in the samples with higher growth temperatures \{Figs. \ref{semafm}(g) and (h)\}. Surface roughness also increases with higher growth temperatures. The mean roughness values for the samples grown at 25$\degree$C, 60$\degree$C, 90$\degree$C, and 120$\degree$C are 0.446 nm, 0.479 nm, 0.626 nm and 1.156 nm, respectively. The elongated Sb formation along the [$\bar1$10] crystalline direction can be attributed to the anisotropic lattice matching of the rhombohedral Sb(104)$_r$ layer on the cubic GaSb(001)$_c$ substrate. More of these structures tend to form as the surface energy increases with the substrate temperature. 

\subsection{Anisotropic electrical transport}
An anisotropy similiar to that seen in surface morphology was also observed in electrical transport. The electrical resistance of  the above four samples was  measured in the square van der Pauw geometry at temperatures as low as 4~K (Fig. \ref{transport12}). Longitudinal resistance \emph{R$_{yy}$} along the direction of the elongated Sb structures, the [$\bar1$10] crystalline orientation, shows lower values compared to \emph{R$_{xx}$} which is in the [110] crystalline orientation. Due to the nature of the van der Pauw geometry, both \emph{R$_{xx}$} and \emph{R$_{yy}$} have contributions of electrical currents flowing in both [110] and [$\bar1$10] directions. We assume \emph{R$_{xx}$} has more contribution from the current along the [110] direction whereas \emph{R$_{yy}$} has more contribution from the current along the [$\bar1$10] direction. The temperature range of interest is below 150~K, where charge carries in the GaSb buffer/substrate freeze, and its resistivity exponentially increases to be several orders of magnitude higher than that of Sb films. Samples grown at 25$\degree$C and 60$\degree$C show similar temperature dependence with metallic behaviors in both \emph{R$_{xx}$} and \emph{R$_{yy}$}, and the ratio of \emph{R$_{xx}$} over \emph{R$_{yy}$} is in the range of 2.1 - 2.5 at 4~K \{Fig. \ref{transport12}(a)\}. The longitudinal resistance of both crystalline orientations becomes more anisotropic as the substrate temperature increases. In  samples grown at 90$\degree$C and 120$\degree$C, the ratio of \emph{R$_{xx}$} over \emph{R$_{yy}$} dramatically increases to be 8 and 545, respectively, at 4~K.  We attribute the anisotropic transport features of the Sb films to the anistropic structure formation. The elongated, wire-like features along the [$\bar1$10] direction result in lower resistance in \emph{R$_{yy}$}. In contrast, electrons moving along the [110] direction see more grain boundaries and curvature on the surface, which result in higher resistance and non-metallic temperature dependence in \emph{R$_{xx}$}.

In addition, longitudinal (\emph{R$_{xx}$} and \emph{R$_{yy}$}) and transverse (\emph{R$_{xy}$}) resistances were measured with respect to the perpendicular magnetic field (\emph{H}) at different temperatures. Figures \ref{transport12}(b) and (c) show  representative longitudinal and transverse curves for the sample grown at 25~$\degree$C. Measurements conducted on the other three samples revealed similar characteristics. The longitudinal resistance as a function of magnetic field (\emph{R} vs \emph{H})  displays a parabolic behavior in both \emph{R$_{xx}$} and \emph{R$_{yy}$} down to 4~K in all four samples. The transverse resistance as a function of magnetic field (\emph{R$_{xy}$} vs \emph{H}) exhibits a linear behavior. p-type carrier density of \emph{n$_{3D}$} = $9.82\times10^{20}$ $cm^{-3}$ and hole mobility of 327.9 $cm^{2}/Vs$ were obtained for the sample grown at 25~$\degree$C. The high carrier density and the metallic temperature dependence of {R$_{xx}$ confirm the semimetallic nature of the Sb films. It is likely that one type of carrier (holes) dominate the transport mechanism, while the contribution from the other carrier (electrons) is negligible.  

\begin{figure}[t!]
    \centering
    \includegraphics[width= \textwidth]{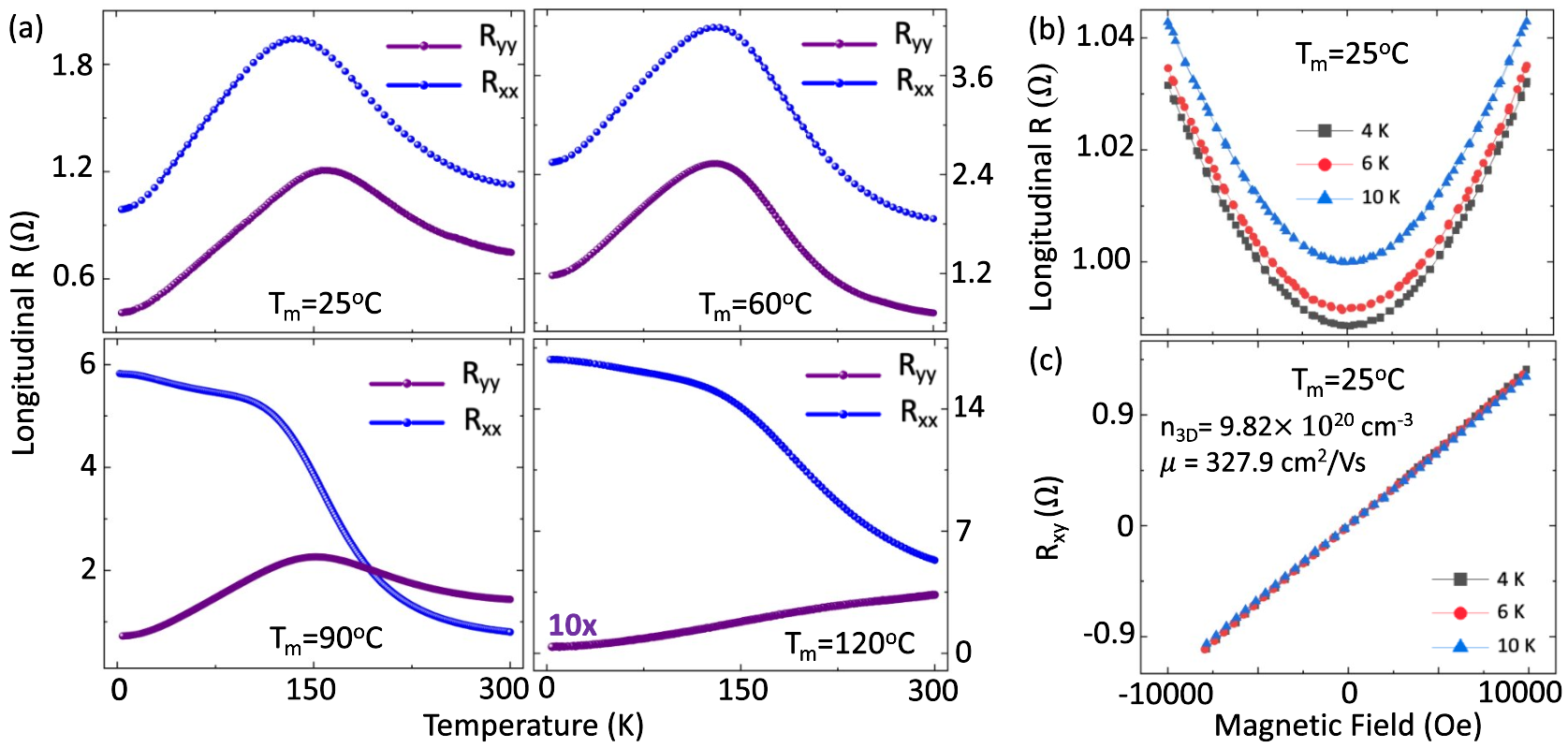}
    \caption{(a) Resistance (\emph{R$_{xx}$} and \emph{R$_{yy}$}) vs temperature curves reveal anisotropic electrical properties in all four samples grown at 25$\degree$C, 60$\degree$C, 90$\degree$C, and 120$\degree$C by using the square van der Pauw geometry in the [110] and [$\bar1$10] directions, respectively. (b) Longitudinal resistance (\emph{R$_{xx}$}) vs magnetic field (\emph{H}) shows a parabolic behavior. (c) Transverse resistance (\emph{R$_{xy}$}) vs magnetic field (\emph{H}) shows a linear behavior at cryogenic temperatures. }  
    \label{transport12}
\end{figure}

\subsection {Anisotropic Sb film growth: surface reconstruction and formation energies}
 
\begin{figure}[t!]
    \centering
    \includegraphics[width=0.8\textwidth]{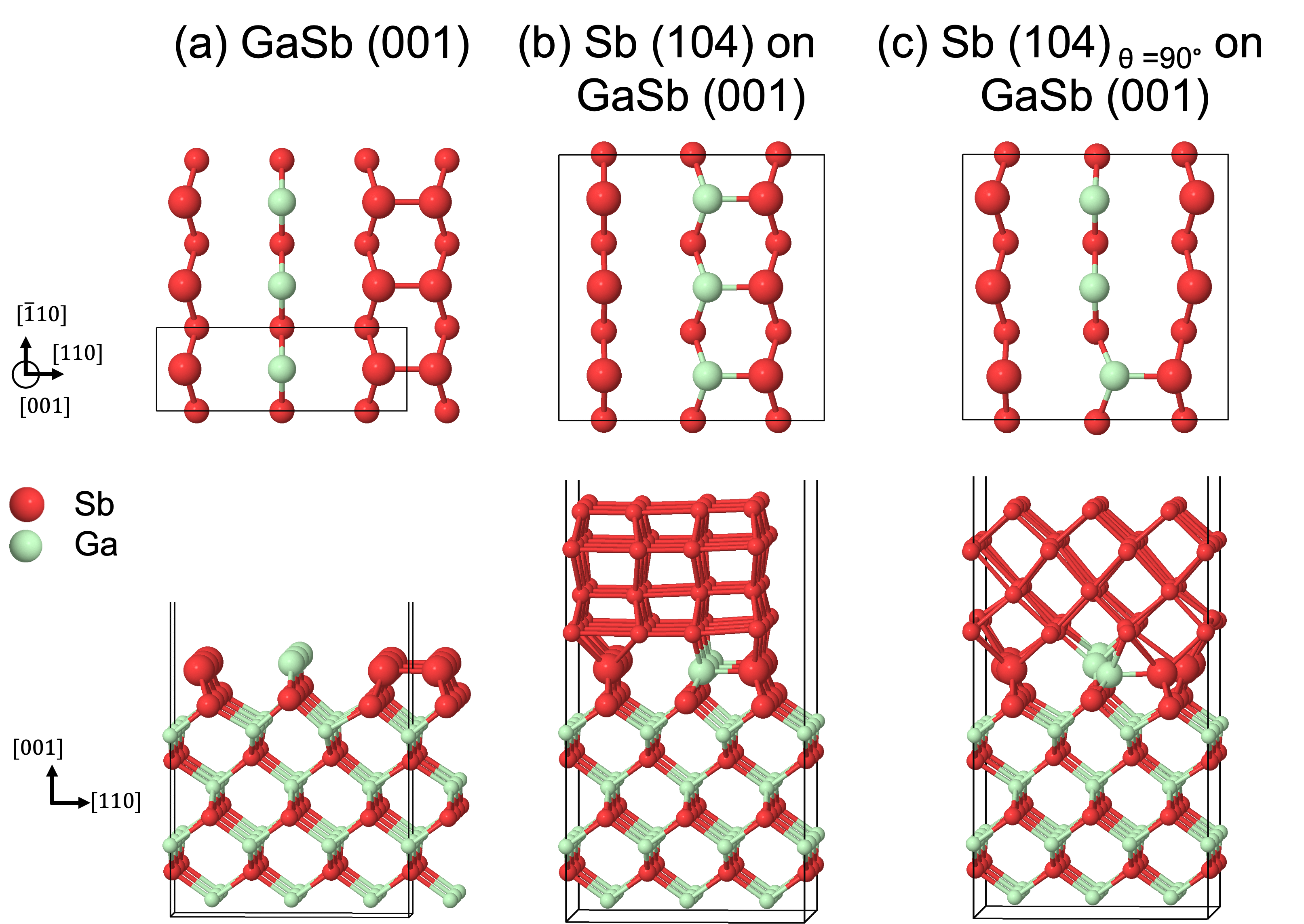}
    \caption{First-principles modeling of Sb films on GaSb substrates. (a) Atomic structures of (1 $\times$ 3) reconstructed cubic GaSb(001). The (104) plane of the rohombohedral Sb film is well aligned with the (001) plane of the GaSb substrate in two different orientations shown in (b) and (c). The formation energy of the Sb(104)$_r$ (b) on the reconstructed substrate (8.34 \textit{meV}/\r{A}$^2$) is found to be lower than that of (c) the 90$\degree$-rotated plane (10.9 \textit{meV}/\r{A}$^2$). The upper panel shows the atomic configurations of the GaSb at the interface with the 2D Sb.}
    \label{GaSb_Sb_modeling}
\end{figure}

To gain further insight into the growth mechanisms of anisotropic Sb thin films, density functional theory (DFT) calculations were performed using the FHI-aims code \cite{Havu2009EfficientFunctions,Knuth2015All-electronOrbitals, Yu2018ELSI:Solvers, Yu2021GPU-accelerationEigenproblems, Huhn2020GPGPUFunctions}, an all-electron code with localized numerical orbitals as the basis, tight basis sets, and the Perdew-Burke-Ernzerhof (PBE) exchange-correlation functional \cite{Perdew1996GeneralizedSimple}. We employed the Hirshfeld scheme for the van der Waals interactions, which are important for the description of the layer interactions and also the interaction between the 2D film and the substrate \cite{Tkatchenko2008AccurateData}. The Broyden-Fletcher-Goldfarb-Shanno (BFGS) \cite{Nocedal1999NumericalOptimization} algorithm was used for atomic relaxations with a condition that the maximum force component is less than $5 \times 10^{-3}$~eV/Å. \\
We conducted a full relaxation of the GaSb unit cell structure and obtained a cubic lattice parameter of 6.125~Å. Using this structure, we generated a $3/\sqrt{2}  \times 1/\sqrt{2} \times 4$ (12.99 $\times$ 4.330 $\times$ 54.52 Å$^3$, ~35Å vacuum) (001)$_c$ slab, which underwent structure relaxation with fixed lattice parameters. Subsequently, a (1 $\times$ 3) surface reconstructed (001)$_c$ structure emerged, as shown in Fig. \ref{GaSb_Sb_modeling}(a), aligning with findings from previous experimental studies\cite{Franklin1990PhotoemissionGaSb100,Sieger1995ReflectionReconstructions}. The 1 $\times$ 3 supercell (12.99 $\times$ 12.99~Å) of the surface-reconstructed GaSb substrate was employed as the substrate for placing a fully relaxed rhombohedral Sb slab - we found the supercell of 2 $\times$ 3 $\times$ 2(104)$_r$ slab with 12.38 x 13.01  Å$^2$ matches well the lattice parameter of the substrate. The GaSb(001)$_c$ substrate has a cubic structure with 4-fold symmetry. The bulk crystal structure along the [110]$_c$ orientation is identical to that along the [$\bar1$10]$_c$ orientation. The formation of elongated Sb structures along the [$\bar{1}10$]$_c$ orientation, but not along the [110]$_c$ orientation, on GaSb(001)$_c$ substrate can be attributed to the surface reconstruction of the GaSb substrate. The anisotropic Sb thin films were grown on the GaSb surface with a $(1 \times 3)$ surface reconstruction, as shown in {Fig. \ref{GaSb_Sb_modeling}(a)}. In this reconstructed surface, atoms on the top layers are distorted to minimize the surface free energy, leading to prominent features along the [$\bar1$10]$_c$ orientation. As a result, the Sb atoms are more likely to align with the Ga and Sb atoms on the reconstructed surface, resulting in the observed elongated Sb structures along the [$\bar1$10]$_c$ orientation.  

We considered two different cases for the incorporation of the (104)$_r$ slab: one in which the lowest Sb atoms of the Sb (104)$_r$ slab are aligned with the top Ga and Sb atoms of the GaSb substrate, and another with the (104)$_r$ slab rotated 90$\degree$. Both structures were fully relaxed, and the stable configurations are shown in Fig.~ \ref{GaSb_Sb_modeling}. The formation energy of the 2D Sb was defined as \(\Delta{E} = (E_{total} - N_{GaSb} \times E_{GaSb}^{3\times1} - N_{Sb} \times E_{Sb}^{bulk})/A\), where $E_{total}$ is the total energy of the system consisting of 2D Sb on the GaSb substrate, $E_{GaSb}^{1\times3}$ is the energy per atom of the $1\times3$-reconstructed GaSb, $E_{Sb}^{bulk}$ is the energy per atom of the rhombohedral Sb bulk, a parent structure of the 2D film, $N_{GaSb}$ and $N_{Sb}$ are the number of atoms in the GaSb substrate and the Sb film, and $A$ is the in-plane area of the supercell of the 2D Sb on GaSb. The formation energy of the Sb(104)$_r$ on the reconstructed substrate (8.34 \textit{meV}/\r{A}$^2$) is found to be lower than that of the 90$\degree$-rotated plane (10.9 \textit{meV}/\r{A}$^2$). This lower formation energy suggests that the observed Sb(104)$_r$ plane is energetically more favorable, further explaining the preferential growth of elongated Sb structures along the [$\bar1$10]$_c$ orientation. 
 
These calculations help to evaluate the energetics and structural stability of different configurations, which ultimately affect the growth behavior of the Sb films on GaSb(001)$_c$ substrates. This preferential growth of Sb structures along the [$\bar{1}10$]$_c$ orientation is further supported by the analysis of the formation energies. The DFT calculations reproduced the experimentally verified reconstruction of the substrate, which is the key to the growth of the Sb(104)$_r$ films with preferred orientation. 

\section{Conclusions}
In summary, we successfully mapped the growth regime of amorphous and crystalline Sb thin films on GaSb(001) surface. We found that there is a transitional region between 250$\degree$C and 150$\degree$C where Sb atoms start to nucleate with diffusion of Ga atoms to form GaSb patches and then amorphous Sb film, confirmed by TEM. By avoiding Sb nucleation across the transitional region during the cooling process, crystalline Sb thin films was coherently grown on GaSb(001) below 120$\degree$C. The crystal structure of the crystalline Sb thin films was found to be rhombohedral with Sb(104)$_r$ plane parallel to the cubic GaSb(001)$_c$ plane. At the interface, atoms of the rhombohedral Sb layer closely align with the GaSb lattice along the [$\bar1$10] zone axis, but not along the [110] zone axis. This anisotropic lattice matching can be attributed to the  (1 $\times$ 3) surface reconstruction of  the GaSb(001) surface, as suggested by our DFT calculations. These calculations provide valuable insights into the formation energy and structural stability of different configurations, which in turn influence the growth behavior of Sb films on the GaSb(001)$_c$ substrates.
The reduced formation energy of Sb(104)$_r$ on the reconstructed substrate further enhances the preferential growth of elongated Sb structures along the [$\bar1$10]$_c$ orientation, leading to streaky/spotty RHEED patterns and anisotropic electronic transport. Such anisotropy is more prominent on the samples grown at higher temperatures. The mean surface roughness of Sb thin film grown at room temperature is 2.5 times smaller than that of Sb thin film grown at 120$\degree$C. The ratio of resistance along [110] direction over the resistance along [$\bar1$10] is two orders of magnitude higher for Sb thin film grown at 120$\degree$C in comparison to the one grown at room temperature. The successful demonstration of epitaxial Sb thin films on cubic GaSb(001) substrates opens a new avenue to embed rhombohedral Sb films on various cubic substrate even with the fact that the cubic Sb phase is unstable. The systematic change in anisotropic features in the Sb thin films suggests optimal growth conditions for further studies and future application using Sb thin films. For topological phases induced by the quantum confinement effect, smooth surface of Sb thin films with minimal anisotropy is preferred to achieve uniform quantum confinement effect. For electrical and thermal transport, the crystalline orientation ([$\bar1$10]  versus [110]) needs to be considered according to the device applications.  
\section*{Acknowledgement}
This work was supported by the Science Alliance at the University of Tennessee, Knoxville, through the Support for Affiliated Research Teams program, by the High-Potential Individuals Global Training Program (Task No. 2021-0-01580) through the Institute of Information and Communications Technology Planning \& Evaluation (IITP) funded by the Republic of Korea’s Ministry of Science and ICT (MSIT) and by the U.S. Department of Energy (DOE), Office of Science, Basic Energy Sciences (BES), Materials Sciences and Engineering Division (MSED) (S. Y., M. B. and A.R.M) and by the
U.S. Department of Energy (DOE), Office of Science, National Quantum Information Science
Research Centers, Quantum Science Center (M.Y.).
This research used resources of the Oak Ridge Leadership Computing Facility (OLCF) and the Compute and Data Environment for Science (CADES) at the Oak Ridge National Laboratory, which are supported by the Office of Science of the U.S. Department of Energy under Contract No. DE-AC05-00OR22725 and of the National Energy Research Scientific Computing Center (NERSC), a U.S. Department of Energy Office of Science User Facility located at Lawrence Berkeley National Laboratory, operated under Contract No. DE-AC02-05CH11231 using NERSC award BES-ERCAP0024568.

\bibliographystyle{apsrev4-1}

\bibliography{references}

\begin{thebibliography}{37}%
\makeatletter
\providecommand \@ifxundefined [1]{%
 \@ifx{#1\undefined}
}%
\providecommand \@ifnum [1]{%
 \ifnum #1\expandafter \@firstoftwo
 \else \expandafter \@secondoftwo
 \fi
}%
\providecommand \@ifx [1]{%
 \ifx #1\expandafter \@firstoftwo
 \else \expandafter \@secondoftwo
 \fi
}%
\providecommand \natexlab [1]{#1}%
\providecommand \enquote  [1]{``#1''}%
\providecommand \bibnamefont  [1]{#1}%
\providecommand \bibfnamefont [1]{#1}%
\providecommand \citenamefont [1]{#1}%
\providecommand \href@noop [0]{\@secondoftwo}%
\providecommand \href [0]{\begingroup \@sanitize@url \@href}%
\providecommand \@href[1]{\@@startlink{#1}\@@href}%
\providecommand \@@href[1]{\endgroup#1\@@endlink}%
\providecommand \@sanitize@url [0]{\catcode `\\12\catcode `\$12\catcode
  `\&12\catcode `\#12\catcode `\^12\catcode `\_12\catcode `\%12\relax}%
\providecommand \@@startlink[1]{}%
\providecommand \@@endlink[0]{}%
\providecommand \url  [0]{\begingroup\@sanitize@url \@url }%
\providecommand \@url [1]{\endgroup\@href {#1}{\urlprefix }}%
\providecommand \urlprefix  [0]{URL }%
\providecommand \Eprint [0]{\href }%
\providecommand \doibase [0]{http://dx.doi.org/}%
\providecommand \selectlanguage [0]{\@gobble}%
\providecommand \bibinfo  [0]{\@secondoftwo}%
\providecommand \bibfield  [0]{\@secondoftwo}%
\providecommand \translation [1]{[#1]}%
\providecommand \BibitemOpen [0]{}%
\providecommand \bibitemStop [0]{}%
\providecommand \bibitemNoStop [0]{.\EOS\space}%
\providecommand \EOS [0]{\spacefactor3000\relax}%
\providecommand \BibitemShut  [1]{\csname bibitem#1\endcsname}%
\let\auto@bib@innerbib\@empty
\bibitem [{\citenamefont {Zhang}\ \emph {et~al.}(2018)\citenamefont {Zhang},
  \citenamefont {Guo}, \citenamefont {Chen}, \citenamefont {Wang},
  \citenamefont {Gao}, \citenamefont {G{\'{o}}~Mez-Herrero}, \citenamefont
  {Ares}, \citenamefont {Lix~Zamora}, \citenamefont {Zhu},\ and\ \citenamefont
  {Zeng}}]{Zhang2018RecentExperiment}%
  \BibitemOpen
  \bibfield  {author} {\bibinfo {author} {\bibfnamefont {S.}~\bibnamefont
  {Zhang}}, \bibinfo {author} {\bibfnamefont {S.}~\bibnamefont {Guo}}, \bibinfo
  {author} {\bibfnamefont {Z.}~\bibnamefont {Chen}}, \bibinfo {author}
  {\bibfnamefont {Y.}~\bibnamefont {Wang}}, \bibinfo {author} {\bibfnamefont
  {H.}~\bibnamefont {Gao}}, \bibinfo {author} {\bibfnamefont {J.}~\bibnamefont
  {G{\'{o}}~Mez-Herrero}}, \bibinfo {author} {\bibfnamefont {P.}~\bibnamefont
  {Ares}}, \bibinfo {author} {\bibfnamefont {F.}~\bibnamefont {Lix~Zamora}},
  \bibinfo {author} {\bibfnamefont {Z.}~\bibnamefont {Zhu}}, \ and\ \bibinfo
  {author} {\bibfnamefont {H.}~\bibnamefont {Zeng}},\ }\href {\doibase
  10.1039/c7cs00125h} {\bibfield  {journal} {\bibinfo  {journal} {Chemical
  Society Reviews}\ }\textbf {\bibinfo {volume} {47}},\ \bibinfo {pages} {982}
  (\bibinfo {year} {2018})}\BibitemShut {NoStop}%
\bibitem [{\citenamefont {Wu}\ and\ \citenamefont
  {Hao}(2020)}]{Wu2020ElectricalLayers}%
  \BibitemOpen
  \bibfield  {author} {\bibinfo {author} {\bibfnamefont {Z.}~\bibnamefont
  {Wu}}\ and\ \bibinfo {author} {\bibfnamefont {J.}~\bibnamefont {Hao}},\
  }\href {\doibase 10.1038/s41699-020-0139-x} {\bibfield  {journal} {\bibinfo
  {journal} {npj 2D Materials and Applications}\ }\textbf {\bibinfo {volume}
  {4}},\ \bibinfo {pages} {4} (\bibinfo {year} {2020})}\BibitemShut {NoStop}%
\bibitem [{\citenamefont {Gui}\ \emph {et~al.}(2019)\citenamefont {Gui},
  \citenamefont {Jin}, \citenamefont {Sun}, \citenamefont {Jiang},\ and\
  \citenamefont {Sun}}]{Gui2019Two-dimensionalApplications}%
  \BibitemOpen
  \bibfield  {author} {\bibinfo {author} {\bibfnamefont {R.}~\bibnamefont
  {Gui}}, \bibinfo {author} {\bibfnamefont {H.}~\bibnamefont {Jin}}, \bibinfo
  {author} {\bibfnamefont {Y.}~\bibnamefont {Sun}}, \bibinfo {author}
  {\bibfnamefont {X.}~\bibnamefont {Jiang}}, \ and\ \bibinfo {author}
  {\bibfnamefont {Z.}~\bibnamefont {Sun}},\ }\href {\doibase
  10.1039/c9ta09582a} {\bibfield  {journal} {\bibinfo  {journal} {Journal of
  Materials Chemistry A}\ }\textbf {\bibinfo {volume} {7}},\ \bibinfo {pages}
  {25712} (\bibinfo {year} {2019})}\BibitemShut {NoStop}%
\bibitem [{\citenamefont {Zhao}\ \emph {et~al.}(2020)\citenamefont {Zhao},
  \citenamefont {Li}, \citenamefont {Hu}, \citenamefont {Wang}, \citenamefont
  {Zhang}, \citenamefont {Lu}, \citenamefont {Ruan},\ and\ \citenamefont
  {Zeng}}]{Chen2020REVIEW2020}%
  \BibitemOpen
  \bibfield  {author} {\bibinfo {author} {\bibfnamefont {A.}~\bibnamefont
  {Zhao}}, \bibinfo {author} {\bibfnamefont {H.}~\bibnamefont {Li}}, \bibinfo
  {author} {\bibfnamefont {X.}~\bibnamefont {Hu}}, \bibinfo {author}
  {\bibfnamefont {C.}~\bibnamefont {Wang}}, \bibinfo {author} {\bibfnamefont
  {H.}~\bibnamefont {Zhang}}, \bibinfo {author} {\bibfnamefont
  {J.}~\bibnamefont {Lu}}, \bibinfo {author} {\bibfnamefont {S.}~\bibnamefont
  {Ruan}}, \ and\ \bibinfo {author} {\bibfnamefont {Y.-J.}\ \bibnamefont
  {Zeng}},\ }\href {\doibase 10.1088/1361-6463/ab810c} {\bibfield  {journal}
  {\bibinfo  {journal} {Journal of Physics D: Applied Physics}\ }\textbf
  {\bibinfo {volume} {53}},\ \bibinfo {pages} {293002} (\bibinfo {year}
  {2020})}\BibitemShut {NoStop}%
\bibitem [{\citenamefont {Zhang}\ \emph {et~al.}(2012)\citenamefont {Zhang},
  \citenamefont {Liu}, \citenamefont {Duan}, \citenamefont {Liu},\ and\
  \citenamefont {Wu}}]{Zhang2012TopologicalEffectb}%
  \BibitemOpen
  \bibfield  {author} {\bibinfo {author} {\bibfnamefont {P.}~\bibnamefont
  {Zhang}}, \bibinfo {author} {\bibfnamefont {Z.}~\bibnamefont {Liu}}, \bibinfo
  {author} {\bibfnamefont {W.}~\bibnamefont {Duan}}, \bibinfo {author}
  {\bibfnamefont {F.}~\bibnamefont {Liu}}, \ and\ \bibinfo {author}
  {\bibfnamefont {J.}~\bibnamefont {Wu}},\ }\href
  {https://doi.org/10.1103/PhysRevB.85.201410} {\bibfield  {journal} {\bibinfo
  {journal} {Physical Review B - Condensed Matter and Materials Physics}\
  }\textbf {\bibinfo {volume} {85}} (\bibinfo {year} {2012})}\BibitemShut
  {NoStop}%
\bibitem [{\citenamefont {Lima}\ and\ \citenamefont
  {Schmidt}(2015)}]{Lima2015TopologicalBilayers}%
  \BibitemOpen
  \bibfield  {author} {\bibinfo {author} {\bibfnamefont {E.~N.}\ \bibnamefont
  {Lima}}\ and\ \bibinfo {author} {\bibfnamefont {T.~M.}\ \bibnamefont
  {Schmidt}},\ }\href {\doibase 10.1103/PhysRevB.91.075432} {\bibfield
  {journal} {\bibinfo  {journal} {Phys. Rev. B}\ }\textbf {\bibinfo {volume}
  {91}},\ \bibinfo {pages} {075432} (\bibinfo {year} {2015})}\BibitemShut
  {NoStop}%
\bibitem [{\citenamefont {Murakami}(2006)}]{Murakami2006QuantumCoupling}%
  \BibitemOpen
  \bibfield  {author} {\bibinfo {author} {\bibfnamefont {S.}~\bibnamefont
  {Murakami}},\ }\href {https://doi.org/10.1103/PhysRevLett.97.236805}
  {\bibfield  {journal} {\bibinfo  {journal} {Physical Review Letters}\
  }\textbf {\bibinfo {volume} {97}} (\bibinfo {year} {2006})}\BibitemShut
  {NoStop}%
\bibitem [{\citenamefont {Wang}\ \emph {et~al.}(2015)\citenamefont {Wang},
  \citenamefont {Chen}, \citenamefont {Liu}, \citenamefont {Wang},
  \citenamefont {Cui}, \citenamefont {Zhang}, \citenamefont {Zhao},\ and\
  \citenamefont {Ji}}]{Wang2015TopologicalAdsorption}%
  \BibitemOpen
  \bibfield  {author} {\bibinfo {author} {\bibfnamefont {D.}~\bibnamefont
  {Wang}}, \bibinfo {author} {\bibfnamefont {L.}~\bibnamefont {Chen}}, \bibinfo
  {author} {\bibfnamefont {H.}~\bibnamefont {Liu}}, \bibinfo {author}
  {\bibfnamefont {X.}~\bibnamefont {Wang}}, \bibinfo {author} {\bibfnamefont
  {G.}~\bibnamefont {Cui}}, \bibinfo {author} {\bibfnamefont {P.}~\bibnamefont
  {Zhang}}, \bibinfo {author} {\bibfnamefont {D.}~\bibnamefont {Zhao}}, \ and\
  \bibinfo {author} {\bibfnamefont {S.}~\bibnamefont {Ji}},\ }\href {\doibase
  10.1039/c4cp04502e} {\bibfield  {journal} {\bibinfo  {journal} {Physical
  Chemistry Chemical Physics}\ }\textbf {\bibinfo {volume} {17}},\ \bibinfo
  {pages} {3577} (\bibinfo {year} {2015})}\BibitemShut {NoStop}%
\bibitem [{\citenamefont {Ares}\ \emph {et~al.}(2017)\citenamefont {Ares},
  \citenamefont {Palacios}, \citenamefont {Abell{\'{a}}n}, \citenamefont
  {G{\'{o}}mez-Herrero},\ and\ \citenamefont
  {Zamora}}]{Ares2017RecentMaterial}%
  \BibitemOpen
  \bibfield  {author} {\bibinfo {author} {\bibfnamefont {P.}~\bibnamefont
  {Ares}}, \bibinfo {author} {\bibfnamefont {J.~J.}\ \bibnamefont {Palacios}},
  \bibinfo {author} {\bibfnamefont {G.}~\bibnamefont {Abell{\'{a}}n}}, \bibinfo
  {author} {\bibfnamefont {J.}~\bibnamefont {G{\'{o}}mez-Herrero}}, \ and\
  \bibinfo {author} {\bibfnamefont {F.}~\bibnamefont {Zamora}},\ }\href
  {\doibase 10.1002/adma.201703771} {\bibfield  {journal} {\bibinfo  {journal}
  {Advanced Materials}\ }\textbf {\bibinfo {volume} {30}},\ \bibinfo {pages}
  {1703771} (\bibinfo {year} {2017})}\BibitemShut {NoStop}%
\bibitem [{\citenamefont {Wang}\ \emph {et~al.}(2019)\citenamefont {Wang},
  \citenamefont {Song},\ and\ \citenamefont
  {Qu}}]{Wang2019Antimonene:Application}%
  \BibitemOpen
  \bibfield  {author} {\bibinfo {author} {\bibfnamefont {X.}~\bibnamefont
  {Wang}}, \bibinfo {author} {\bibfnamefont {J.}~\bibnamefont {Song}}, \ and\
  \bibinfo {author} {\bibfnamefont {J.}~\bibnamefont {Qu}},\ }\href {\doibase
  https://doi.org/10.1002/anie.201808302} {\bibfield  {journal} {\bibinfo
  {journal} {Angewandte Chemie International Edition}\ }\textbf {\bibinfo
  {volume} {58}},\ \bibinfo {pages} {1574} (\bibinfo {year}
  {2019})}\BibitemShut {NoStop}%
\bibitem [{\citenamefont {Liu}\ \emph {et~al.}(2020)\citenamefont {Liu},
  \citenamefont {Zhang},\ and\ \citenamefont
  {Yang}}]{Liu2020Antimonene:Material}%
  \BibitemOpen
  \bibfield  {author} {\bibinfo {author} {\bibfnamefont {S.}~\bibnamefont
  {Liu}}, \bibinfo {author} {\bibfnamefont {T.}~\bibnamefont {Zhang}}, \ and\
  \bibinfo {author} {\bibfnamefont {S.}~\bibnamefont {Yang}},\ }\enquote
  {\bibinfo {title} {Antimonene: A potential 2d material},}\ in\ \href
  {\doibase https://doi.org/10.1002/9781119655275.ch2} {\emph {\bibinfo
  {booktitle} {Monoelements}}}\ (\bibinfo  {publisher} {John Wiley I\& Sons,
  Ltd},\ \bibinfo {year} {2020})\ Chap.~\bibinfo {chapter} {2}, pp.\ \bibinfo
  {pages} {27--56}\BibitemShut {NoStop}%
\bibitem [{\citenamefont {Xue}\ and\ \citenamefont
  {Li}(2021)}]{Xue2021RecentGrowth}%
  \BibitemOpen
  \bibfield  {author} {\bibinfo {author} {\bibfnamefont {C.-L.}\ \bibnamefont
  {Xue}}\ and\ \bibinfo {author} {\bibfnamefont {S.-C.}\ \bibnamefont {Li}},\
  }\href {\doibase 10.35848/1347-4065/abf74e} {\bibfield  {journal} {\bibinfo
  {journal} {Japanese Journal of Applied Physics}\ }\textbf {\bibinfo {volume}
  {60}},\ \bibinfo {pages} {SE0805} (\bibinfo {year} {2021})}\BibitemShut
  {NoStop}%
\bibitem [{\citenamefont {Brahlek}\ \emph {et~al.}(2020)\citenamefont
  {Brahlek}, \citenamefont {Lapano},\ and\ \citenamefont
  {Lee}}]{Brahlek2020TopologicalEpitaxy}%
  \BibitemOpen
  \bibfield  {author} {\bibinfo {author} {\bibfnamefont {M.}~\bibnamefont
  {Brahlek}}, \bibinfo {author} {\bibfnamefont {J.}~\bibnamefont {Lapano}}, \
  and\ \bibinfo {author} {\bibfnamefont {J.~S.}\ \bibnamefont {Lee}},\ }\href
  {\doibase 10.1063/5.0022948} {\bibfield  {journal} {\bibinfo  {journal}
  {Journal of Applied Physics}\ }\textbf {\bibinfo {volume} {128}},\ \bibinfo
  {pages} {210902} (\bibinfo {year} {2020})}\BibitemShut {NoStop}%
\bibitem [{\citenamefont {Dumas}\ \emph {et~al.}(1992)\citenamefont {Dumas},
  \citenamefont {Nouaoura}, \citenamefont {Bertru}, \citenamefont
  {Lassabat{\`{e}}re}, \citenamefont {Chen},\ and\ \citenamefont
  {Kahn}}]{Dumas1992Sb-cappingGaSb100}%
  \BibitemOpen
  \bibfield  {author} {\bibinfo {author} {\bibfnamefont {M.}~\bibnamefont
  {Dumas}}, \bibinfo {author} {\bibfnamefont {M.}~\bibnamefont {Nouaoura}},
  \bibinfo {author} {\bibfnamefont {N.}~\bibnamefont {Bertru}}, \bibinfo
  {author} {\bibfnamefont {L.}~\bibnamefont {Lassabat{\`{e}}re}}, \bibinfo
  {author} {\bibfnamefont {W.}~\bibnamefont {Chen}}, \ and\ \bibinfo {author}
  {\bibfnamefont {A.}~\bibnamefont {Kahn}},\ }\href {\doibase
  10.1016/0039-6028(92)90114-L} {\bibfield  {journal} {\bibinfo  {journal}
  {Surface Science}\ }\textbf {\bibinfo {volume} {262}},\ \bibinfo {pages}
  {L91} (\bibinfo {year} {1992})}\BibitemShut {NoStop}%
\bibitem [{\citenamefont {Clark}\ \emph {et~al.}(1995)\citenamefont {Clark},
  \citenamefont {Cairns}, \citenamefont {Wilks}, \citenamefont {Williams},
  \citenamefont {Johnson},\ and\ \citenamefont
  {Whitehouse}}]{Clark1995AntimonyInAlSb100}%
  \BibitemOpen
  \bibfield  {author} {\bibinfo {author} {\bibfnamefont {S.~A.}\ \bibnamefont
  {Clark}}, \bibinfo {author} {\bibfnamefont {J.~W.}\ \bibnamefont {Cairns}},
  \bibinfo {author} {\bibfnamefont {S.~P.}\ \bibnamefont {Wilks}}, \bibinfo
  {author} {\bibfnamefont {R.~H.}\ \bibnamefont {Williams}}, \bibinfo {author}
  {\bibfnamefont {A.~D.}\ \bibnamefont {Johnson}}, \ and\ \bibinfo {author}
  {\bibfnamefont {C.~R.}\ \bibnamefont {Whitehouse}},\ }\href {\doibase
  10.1016/0039-6028(95)00503-X} {\bibfield  {journal} {\bibinfo  {journal}
  {Surface Science}\ }\textbf {\bibinfo {volume} {336}},\ \bibinfo {pages}
  {193} (\bibinfo {year} {1995})}\BibitemShut {NoStop}%
\bibitem [{\citenamefont {Cheng}\ \emph {et~al.}(1984)\citenamefont {Cheng},
  \citenamefont {Zhang},\ and\ \citenamefont
  {Milnes}}]{Cheng1984Schottky-barrierArsenide}%
  \BibitemOpen
  \bibfield  {author} {\bibinfo {author} {\bibfnamefont {H.}~\bibnamefont
  {Cheng}}, \bibinfo {author} {\bibfnamefont {X.~J.}\ \bibnamefont {Zhang}}, \
  and\ \bibinfo {author} {\bibfnamefont {A.~G.}\ \bibnamefont {Milnes}},\
  }\href {\doibase 10.1016/0038-1101(84)90052-2} {\bibfield  {journal}
  {\bibinfo  {journal} {Solid-State Electronics}\ }\textbf {\bibinfo {volume}
  {27}},\ \bibinfo {pages} {1117} (\bibinfo {year} {1984})}\BibitemShut
  {NoStop}%
\bibitem [{\citenamefont {Kwok}\ \emph {et~al.}(1987)\citenamefont {Kwok},
  \citenamefont {Lam},\ and\ \citenamefont {Wong}}]{Kwok1987SchottkySilicon}%
  \BibitemOpen
  \bibfield  {author} {\bibinfo {author} {\bibfnamefont {H.~L.}\ \bibnamefont
  {Kwok}}, \bibinfo {author} {\bibfnamefont {Y.~W.}\ \bibnamefont {Lam}}, \
  and\ \bibinfo {author} {\bibfnamefont {S.~P.}\ \bibnamefont {Wong}},\ }\href
  {\doibase 10.1088/0268-1242/2/5/007} {\bibfield  {journal} {\bibinfo
  {journal} {Semiconductor Science and Technology}\ }\textbf {\bibinfo {volume}
  {2}},\ \bibinfo {pages} {288} (\bibinfo {year} {1987})}\BibitemShut {NoStop}%
\bibitem [{\citenamefont {Gaspe}\ \emph {et~al.}(2013)\citenamefont {Gaspe},
  \citenamefont {Cairns}, \citenamefont {Lei}, \citenamefont {Wickramasinghe},
  \citenamefont {Mishima}, \citenamefont {Keay}, \citenamefont {Murphy},\ and\
  \citenamefont {Santos}}]{Gaspe2013EpitaxialWells}%
  \BibitemOpen
  \bibfield  {author} {\bibinfo {author} {\bibfnamefont {C.~K.}\ \bibnamefont
  {Gaspe}}, \bibinfo {author} {\bibfnamefont {S.}~\bibnamefont {Cairns}},
  \bibinfo {author} {\bibfnamefont {L.}~\bibnamefont {Lei}}, \bibinfo {author}
  {\bibfnamefont {K.~S.}\ \bibnamefont {Wickramasinghe}}, \bibinfo {author}
  {\bibfnamefont {T.~D.}\ \bibnamefont {Mishima}}, \bibinfo {author}
  {\bibfnamefont {J.~C.}\ \bibnamefont {Keay}}, \bibinfo {author}
  {\bibfnamefont {S.~Q.}\ \bibnamefont {Murphy}}, \ and\ \bibinfo {author}
  {\bibfnamefont {M.~B.}\ \bibnamefont {Santos}},\ }\href {\doibase
  10.1116/1.4802212} {\bibfield  {journal} {\bibinfo  {journal} {Journal of
  Vacuum Science {\&} Technology B, Nanotechnology and Microelectronics:
  Materials, Processing, Measurement, and Phenomena}\ }\textbf {\bibinfo
  {volume} {31}},\ \bibinfo {pages} {03C129} (\bibinfo {year}
  {2013})}\BibitemShut {NoStop}%
\bibitem [{\citenamefont {Sun}\ \emph {et~al.}(2018)\citenamefont {Sun},
  \citenamefont {Lu}, \citenamefont {Xiang}, \citenamefont {Wang},
  \citenamefont {Shi}, \citenamefont {Wang}, \citenamefont {Washington},\ and\
  \citenamefont {Lu}}]{Sun2018VanGraphene}%
  \BibitemOpen
  \bibfield  {author} {\bibinfo {author} {\bibfnamefont {X.}~\bibnamefont
  {Sun}}, \bibinfo {author} {\bibfnamefont {Z.}~\bibnamefont {Lu}}, \bibinfo
  {author} {\bibfnamefont {Y.}~\bibnamefont {Xiang}}, \bibinfo {author}
  {\bibfnamefont {Y.}~\bibnamefont {Wang}}, \bibinfo {author} {\bibfnamefont
  {J.}~\bibnamefont {Shi}}, \bibinfo {author} {\bibfnamefont {G.~C.}\
  \bibnamefont {Wang}}, \bibinfo {author} {\bibfnamefont {M.~A.}\ \bibnamefont
  {Washington}}, \ and\ \bibinfo {author} {\bibfnamefont {T.~M.}\ \bibnamefont
  {Lu}},\ }\href {\doibase 10.1021/acsnano.8b02374} {\bibfield  {journal}
  {\bibinfo  {journal} {ACS Nano}\ }\textbf {\bibinfo {volume} {12}},\ \bibinfo
  {pages} {6100} (\bibinfo {year} {2018})}\BibitemShut {NoStop}%
\bibitem [{\citenamefont {Niu}\ \emph {et~al.}(2019)\citenamefont {Niu},
  \citenamefont {Zhou}, \citenamefont {Zhou}, \citenamefont {Hu}, \citenamefont
  {Zhang}, \citenamefont {Zhang}, \citenamefont {Zhou}, \citenamefont {Fuchs},\
  and\ \citenamefont {Zeng}}]{Niu2019ModulatingDesign}%
  \BibitemOpen
  \bibfield  {author} {\bibinfo {author} {\bibfnamefont {T.}~\bibnamefont
  {Niu}}, \bibinfo {author} {\bibfnamefont {W.}~\bibnamefont {Zhou}}, \bibinfo
  {author} {\bibfnamefont {D.}~\bibnamefont {Zhou}}, \bibinfo {author}
  {\bibfnamefont {X.}~\bibnamefont {Hu}}, \bibinfo {author} {\bibfnamefont
  {S.}~\bibnamefont {Zhang}}, \bibinfo {author} {\bibfnamefont
  {K.}~\bibnamefont {Zhang}}, \bibinfo {author} {\bibfnamefont
  {M.}~\bibnamefont {Zhou}}, \bibinfo {author} {\bibfnamefont {H.}~\bibnamefont
  {Fuchs}}, \ and\ \bibinfo {author} {\bibfnamefont {H.}~\bibnamefont {Zeng}},\
  }\href {https://doi.org/10.1002/adma.201902606} {\bibfield  {journal}
  {\bibinfo  {journal} {Advanced Materials}\ }\textbf {\bibinfo {volume} {31}}
  (\bibinfo {year} {2019})}\BibitemShut {NoStop}%
\bibitem [{\citenamefont {Dura}\ \emph {et~al.}(1995)\citenamefont {Dura},
  \citenamefont {Vigliante}, \citenamefont {Golding},\ and\ \citenamefont
  {Moss}}]{Dura1995EpitaxialHeteroepitaxy}%
  \BibitemOpen
  \bibfield  {author} {\bibinfo {author} {\bibfnamefont {J.~A.}\ \bibnamefont
  {Dura}}, \bibinfo {author} {\bibfnamefont {A.}~\bibnamefont {Vigliante}},
  \bibinfo {author} {\bibfnamefont {T.~D.}\ \bibnamefont {Golding}}, \ and\
  \bibinfo {author} {\bibfnamefont {S.~C.}\ \bibnamefont {Moss}},\ }\href
  {\doibase 10.1063/1.359373} {\bibfield  {journal} {\bibinfo  {journal}
  {Journal of Applied Physics}\ }\textbf {\bibinfo {volume} {77}},\ \bibinfo
  {pages} {21} (\bibinfo {year} {1995})}\BibitemShut {NoStop}%
\bibitem [{\citenamefont {Donohue}(1974)}]{Donohue1974TheElements.}%
  \BibitemOpen
  \bibfield  {author} {\bibinfo {author} {\bibfnamefont {J.}~\bibnamefont
  {Donohue}},\ }\href@noop {} {\emph {\bibinfo {title} {{The structures of the
  elements.}}}}\ (\bibinfo  {publisher} {Wiley},\ \bibinfo {address} {New
  York},\ \bibinfo {year} {1974})\BibitemShut {NoStop}%
\bibitem [{\citenamefont {Hasan}\ and\ \citenamefont
  {Kane}(2010)}]{Hasan2010Colloquium:Insulators}%
  \BibitemOpen
  \bibfield  {author} {\bibinfo {author} {\bibfnamefont {M.~Z.}\ \bibnamefont
  {Hasan}}\ and\ \bibinfo {author} {\bibfnamefont {C.~L.}\ \bibnamefont
  {Kane}},\ }\href {\doibase 10.1103/RevModPhys.82.3045} {\bibfield  {journal}
  {\bibinfo  {journal} {Reviews of Modern Physics}\ }\textbf {\bibinfo {volume}
  {82}},\ \bibinfo {pages} {3045} (\bibinfo {year} {2010})}\BibitemShut
  {NoStop}%
\bibitem [{\citenamefont {Khang}\ \emph {et~al.}(2018)\citenamefont {Khang},
  \citenamefont {Ueda},\ and\ \citenamefont {Hai}}]{HuynhASwitching}%
  \BibitemOpen
  \bibfield  {author} {\bibinfo {author} {\bibfnamefont {N.~H.~D.}\
  \bibnamefont {Khang}}, \bibinfo {author} {\bibfnamefont {Y.}~\bibnamefont
  {Ueda}}, \ and\ \bibinfo {author} {\bibfnamefont {P.~N.}\ \bibnamefont
  {Hai}},\ }\href {https://doi.org/10.1038/s41563-018-0137-y} {\bibfield
  {journal} {\bibinfo  {journal} {Nature materials}\ }\textbf {\bibinfo
  {volume} {17}},\ \bibinfo {pages} {808} (\bibinfo {year} {2018})}\BibitemShut
  {NoStop}%
\bibitem [{\citenamefont {Mellnik}\ \emph {et~al.}(2014)\citenamefont
  {Mellnik}, \citenamefont {Lee}, \citenamefont {Richardella}, \citenamefont
  {Grab}, \citenamefont {Mintun}, \citenamefont {Fischer}, \citenamefont
  {Vaezi}, \citenamefont {Manchon}, \citenamefont {Kim}, \citenamefont
  {Samarth},\ and\ \citenamefont {Ralph}}]{Mellnik2014Spin-transferInsulator}%
  \BibitemOpen
  \bibfield  {author} {\bibinfo {author} {\bibfnamefont {A.~R.}\ \bibnamefont
  {Mellnik}}, \bibinfo {author} {\bibfnamefont {J.~S.}\ \bibnamefont {Lee}},
  \bibinfo {author} {\bibfnamefont {A.}~\bibnamefont {Richardella}}, \bibinfo
  {author} {\bibfnamefont {J.~L.}\ \bibnamefont {Grab}}, \bibinfo {author}
  {\bibfnamefont {P.~J.}\ \bibnamefont {Mintun}}, \bibinfo {author}
  {\bibfnamefont {M.~H.}\ \bibnamefont {Fischer}}, \bibinfo {author}
  {\bibfnamefont {A.}~\bibnamefont {Vaezi}}, \bibinfo {author} {\bibfnamefont
  {A.}~\bibnamefont {Manchon}}, \bibinfo {author} {\bibfnamefont {E.-A.}\
  \bibnamefont {Kim}}, \bibinfo {author} {\bibfnamefont {N.}~\bibnamefont
  {Samarth}}, \ and\ \bibinfo {author} {\bibfnamefont {.~D.~C.}\ \bibnamefont
  {Ralph}},\ }\href {http://doi.org/10.1038/nature13534} {\bibfield  {journal}
  {\bibinfo  {journal} {Nature}\ }\textbf {\bibinfo {volume} {511}},\ \bibinfo
  {pages} {449} (\bibinfo {year} {2014})}\BibitemShut {NoStop}%
\bibitem [{\citenamefont {Bracker}\ \emph {et~al.}(2000)\citenamefont
  {Bracker}, \citenamefont {Yang}, \citenamefont {Bennett}, \citenamefont
  {Culbertson},\ and\ \citenamefont {Moore}}]{Bracker2000SurfaceGaSb}%
  \BibitemOpen
  \bibfield  {author} {\bibinfo {author} {\bibfnamefont {A.}~\bibnamefont
  {Bracker}}, \bibinfo {author} {\bibfnamefont {M.}~\bibnamefont {Yang}},
  \bibinfo {author} {\bibfnamefont {B.}~\bibnamefont {Bennett}}, \bibinfo
  {author} {\bibfnamefont {J.}~\bibnamefont {Culbertson}}, \ and\ \bibinfo
  {author} {\bibfnamefont {W.}~\bibnamefont {Moore}},\ }\href
  {https://doi.org/10.1016/S0022-0248(00)00871-X} {\bibfield  {journal}
  {\bibinfo  {journal} {Journal of crystal growth}\ }\textbf {\bibinfo {volume}
  {220}},\ \bibinfo {pages} {384} (\bibinfo {year} {2000})}\BibitemShut
  {NoStop}%
\bibitem [{\citenamefont {Nouaoura}\ \emph {et~al.}(1997)\citenamefont
  {Nouaoura}, \citenamefont {Da~Silva}, \citenamefont {Bertru}, \citenamefont
  {Rouanet}, \citenamefont {Tahraoui}, \citenamefont {Oueini}, \citenamefont
  {Bonnet},\ and\ \citenamefont {Lassabatere}}]{Nouaoura1997ModificationFlux}%
  \BibitemOpen
  \bibfield  {author} {\bibinfo {author} {\bibfnamefont {M.}~\bibnamefont
  {Nouaoura}}, \bibinfo {author} {\bibfnamefont {F.~W.}\ \bibnamefont
  {Da~Silva}}, \bibinfo {author} {\bibfnamefont {N.}~\bibnamefont {Bertru}},
  \bibinfo {author} {\bibfnamefont {M.}~\bibnamefont {Rouanet}}, \bibinfo
  {author} {\bibfnamefont {A.}~\bibnamefont {Tahraoui}}, \bibinfo {author}
  {\bibfnamefont {W.}~\bibnamefont {Oueini}}, \bibinfo {author} {\bibfnamefont
  {J.}~\bibnamefont {Bonnet}}, \ and\ \bibinfo {author} {\bibfnamefont
  {L.}~\bibnamefont {Lassabatere}},\ }\href {\doibase
  10.1016/S0022-0248(96)00741-5} {\bibfield  {journal} {\bibinfo  {journal}
  {Journal of Crystal Growth}\ }\textbf {\bibinfo {volume} {172}},\ \bibinfo
  {pages} {37} (\bibinfo {year} {1997})}\BibitemShut {NoStop}%
\bibitem [{\citenamefont {Havu}\ \emph {et~al.}(2009)\citenamefont {Havu},
  \citenamefont {Blum}, \citenamefont {Havu},\ and\ \citenamefont
  {Scheffler}}]{Havu2009EfficientFunctions}%
  \BibitemOpen
  \bibfield  {author} {\bibinfo {author} {\bibfnamefont {V.}~\bibnamefont
  {Havu}}, \bibinfo {author} {\bibfnamefont {V.}~\bibnamefont {Blum}}, \bibinfo
  {author} {\bibfnamefont {P.}~\bibnamefont {Havu}}, \ and\ \bibinfo {author}
  {\bibfnamefont {M.}~\bibnamefont {Scheffler}},\ }\href {\doibase
  https://doi.org/10.1016/j.jcp.2009.08.008} {\bibfield  {journal} {\bibinfo
  {journal} {Journal of Computational Physics}\ }\textbf {\bibinfo {volume}
  {228}},\ \bibinfo {pages} {8367} (\bibinfo {year} {2009})}\BibitemShut
  {NoStop}%
\bibitem [{\citenamefont {Knuth}\ \emph {et~al.}(2015)\citenamefont {Knuth},
  \citenamefont {Carbogno}, \citenamefont {Atalla}, \citenamefont {Blum},\ and\
  \citenamefont {Scheffler}}]{Knuth2015All-electronOrbitals}%
  \BibitemOpen
  \bibfield  {author} {\bibinfo {author} {\bibfnamefont {F.}~\bibnamefont
  {Knuth}}, \bibinfo {author} {\bibfnamefont {C.}~\bibnamefont {Carbogno}},
  \bibinfo {author} {\bibfnamefont {V.}~\bibnamefont {Atalla}}, \bibinfo
  {author} {\bibfnamefont {V.}~\bibnamefont {Blum}}, \ and\ \bibinfo {author}
  {\bibfnamefont {M.}~\bibnamefont {Scheffler}},\ }\href {\doibase
  10.1016/J.CPC.2015.01.003} {\bibfield  {journal} {\bibinfo  {journal}
  {Computer Physics Communications}\ }\textbf {\bibinfo {volume} {190}},\
  \bibinfo {pages} {33} (\bibinfo {year} {2015})}\BibitemShut {NoStop}%
\bibitem [{\citenamefont {Yu}\ \emph {et~al.}(2018)\citenamefont {Yu},
  \citenamefont {Corsetti}, \citenamefont {Garc{\'{i}}a}, \citenamefont {Huhn},
  \citenamefont {Jacquelin}, \citenamefont {Jia}, \citenamefont {Lange},
  \citenamefont {Lin}, \citenamefont {Lu}, \citenamefont {Mi}, \citenamefont
  {Seifitokaldani}, \citenamefont {V{\'{a}}zquez-Mayagoitia}, \citenamefont
  {Yang}, \citenamefont {Yang},\ and\ \citenamefont
  {Blum}}]{Yu2018ELSI:Solvers}%
  \BibitemOpen
  \bibfield  {author} {\bibinfo {author} {\bibfnamefont {V.~W.~z.}\
  \bibnamefont {Yu}}, \bibinfo {author} {\bibfnamefont {F.}~\bibnamefont
  {Corsetti}}, \bibinfo {author} {\bibfnamefont {A.}~\bibnamefont
  {Garc{\'{i}}a}}, \bibinfo {author} {\bibfnamefont {W.~P.}\ \bibnamefont
  {Huhn}}, \bibinfo {author} {\bibfnamefont {M.}~\bibnamefont {Jacquelin}},
  \bibinfo {author} {\bibfnamefont {W.}~\bibnamefont {Jia}}, \bibinfo {author}
  {\bibfnamefont {B.}~\bibnamefont {Lange}}, \bibinfo {author} {\bibfnamefont
  {L.}~\bibnamefont {Lin}}, \bibinfo {author} {\bibfnamefont {J.}~\bibnamefont
  {Lu}}, \bibinfo {author} {\bibfnamefont {W.}~\bibnamefont {Mi}}, \bibinfo
  {author} {\bibfnamefont {A.}~\bibnamefont {Seifitokaldani}}, \bibinfo
  {author} {\bibfnamefont {Ã.}~\bibnamefont {V{\'{a}}zquez-Mayagoitia}},
  \bibinfo {author} {\bibfnamefont {C.}~\bibnamefont {Yang}}, \bibinfo {author}
  {\bibfnamefont {H.}~\bibnamefont {Yang}}, \ and\ \bibinfo {author}
  {\bibfnamefont {V.}~\bibnamefont {Blum}},\ }\href {\doibase
  10.1016/J.CPC.2017.09.007} {\bibfield  {journal} {\bibinfo  {journal}
  {Computer Physics Communications}\ }\textbf {\bibinfo {volume} {222}},\
  \bibinfo {pages} {267} (\bibinfo {year} {2018})}\BibitemShut {NoStop}%
\bibitem [{\citenamefont {Yu}\ \emph {et~al.}(2021)\citenamefont {Yu},
  \citenamefont {Moussa}, \citenamefont {Kůs}, \citenamefont {Marek},
  \citenamefont {Messmer}, \citenamefont {Yoon}, \citenamefont {Lederer},\ and\
  \citenamefont {Blum}}]{Yu2021GPU-accelerationEigenproblems}%
  \BibitemOpen
  \bibfield  {author} {\bibinfo {author} {\bibfnamefont {V.~W.~z.}\
  \bibnamefont {Yu}}, \bibinfo {author} {\bibfnamefont {J.}~\bibnamefont
  {Moussa}}, \bibinfo {author} {\bibfnamefont {P.}~\bibnamefont {Kůs}},
  \bibinfo {author} {\bibfnamefont {A.}~\bibnamefont {Marek}}, \bibinfo
  {author} {\bibfnamefont {P.}~\bibnamefont {Messmer}}, \bibinfo {author}
  {\bibfnamefont {M.}~\bibnamefont {Yoon}}, \bibinfo {author} {\bibfnamefont
  {H.}~\bibnamefont {Lederer}}, \ and\ \bibinfo {author} {\bibfnamefont
  {V.}~\bibnamefont {Blum}},\ }\href {\doibase 10.1016/J.CPC.2020.107808}
  {\bibfield  {journal} {\bibinfo  {journal} {Computer Physics Communications}\
  }\textbf {\bibinfo {volume} {262}},\ \bibinfo {pages} {107808} (\bibinfo
  {year} {2021})}\BibitemShut {NoStop}%
\bibitem [{\citenamefont {Huhn}\ \emph {et~al.}(2020)\citenamefont {Huhn},
  \citenamefont {Lange}, \citenamefont {zhe Yu}, \citenamefont {Yoon},\ and\
  \citenamefont {Blum}}]{Huhn2020GPGPUFunctions}%
  \BibitemOpen
  \bibfield  {author} {\bibinfo {author} {\bibfnamefont {W.~P.}\ \bibnamefont
  {Huhn}}, \bibinfo {author} {\bibfnamefont {B.}~\bibnamefont {Lange}},
  \bibinfo {author} {\bibfnamefont {V.~W.}\ \bibnamefont {zhe Yu}}, \bibinfo
  {author} {\bibfnamefont {M.}~\bibnamefont {Yoon}}, \ and\ \bibinfo {author}
  {\bibfnamefont {V.}~\bibnamefont {Blum}},\ }\href
  {https://doi.org/10.1016/j.cpc.2020.107314} {\bibfield  {journal} {\bibinfo
  {journal} {Computer Physics Communications}\ }\textbf {\bibinfo {volume}
  {254}},\ \bibinfo {pages} {107314} (\bibinfo {year} {2020})}\BibitemShut
  {NoStop}%
\bibitem [{\citenamefont {Perdew}\ \emph {et~al.}(1996)\citenamefont {Perdew},
  \citenamefont {Burke},\ and\ \citenamefont
  {Ernzerhof}}]{Perdew1996GeneralizedSimple}%
  \BibitemOpen
  \bibfield  {author} {\bibinfo {author} {\bibfnamefont {J.~P.}\ \bibnamefont
  {Perdew}}, \bibinfo {author} {\bibfnamefont {K.}~\bibnamefont {Burke}}, \
  and\ \bibinfo {author} {\bibfnamefont {M.}~\bibnamefont {Ernzerhof}},\ }\href
  {\doibase 10.1103/PhysRevLett.77.3865} {\bibfield  {journal} {\bibinfo
  {journal} {Physical Review Letters}\ }\textbf {\bibinfo {volume} {77}},\
  \bibinfo {pages} {3865} (\bibinfo {year} {1996})}\BibitemShut {NoStop}%
\bibitem [{\citenamefont {Tkatchenko}\ and\ \citenamefont
  {Scheffler}(2009)}]{Tkatchenko2008AccurateData}%
  \BibitemOpen
  \bibfield  {author} {\bibinfo {author} {\bibfnamefont {A.}~\bibnamefont
  {Tkatchenko}}\ and\ \bibinfo {author} {\bibfnamefont {M.}~\bibnamefont
  {Scheffler}},\ }\href
  {https://link.aps.org/doi/10.1103/PhysRevLett.102.073005} {\bibfield
  {journal} {\bibinfo  {journal} {Physical Review Letters}\ }\textbf {\bibinfo
  {volume} {102}},\ \bibinfo {pages} {073005} (\bibinfo {year}
  {2009})}\BibitemShut {NoStop}%
\bibitem [{\citenamefont {Nocedal}\ and\ \citenamefont
  {Wright}(1999)}]{Nocedal1999NumericalOptimization}%
  \BibitemOpen
  \bibfield  {author} {\bibinfo {author} {\bibfnamefont {J.}~\bibnamefont
  {Nocedal}}\ and\ \bibinfo {author} {\bibfnamefont {S.~J.}\ \bibnamefont
  {Wright}},\ }\href@noop {} {\emph {\bibinfo {title} {{Numerical
  optimization}}}}\ (\bibinfo  {publisher} {Springer},\ \bibinfo {year}
  {1999})\BibitemShut {NoStop}%
\bibitem [{\citenamefont {Franklin}\ \emph {et~al.}(1990)\citenamefont
  {Franklin}, \citenamefont {Rich}, \citenamefont {Samsavar}, \citenamefont
  {Hirschorn}, \citenamefont {Leibsle}, \citenamefont {Miller},\ and\
  \citenamefont {Chiang}}]{Franklin1990PhotoemissionGaSb100}%
  \BibitemOpen
  \bibfield  {author} {\bibinfo {author} {\bibfnamefont {G.~E.}\ \bibnamefont
  {Franklin}}, \bibinfo {author} {\bibfnamefont {D.~H.}\ \bibnamefont {Rich}},
  \bibinfo {author} {\bibfnamefont {A.}~\bibnamefont {Samsavar}}, \bibinfo
  {author} {\bibfnamefont {E.~S.}\ \bibnamefont {Hirschorn}}, \bibinfo {author}
  {\bibfnamefont {F.~M.}\ \bibnamefont {Leibsle}}, \bibinfo {author}
  {\bibfnamefont {T.}~\bibnamefont {Miller}}, \ and\ \bibinfo {author}
  {\bibfnamefont {T.~C.}\ \bibnamefont {Chiang}},\ }\href {\doibase
  10.1103/PhysRevB.41.12619} {\bibfield  {journal} {\bibinfo  {journal}
  {Physical Review B}\ }\textbf {\bibinfo {volume} {41}},\ \bibinfo {pages}
  {12619} (\bibinfo {year} {1990})}\BibitemShut {NoStop}%
\bibitem [{\citenamefont {Sieger}\ \emph {et~al.}(1995)\citenamefont {Sieger},
  \citenamefont {Miller},\ and\ \citenamefont
  {Chiang}}]{Sieger1995ReflectionReconstructions}%
  \BibitemOpen
  \bibfield  {author} {\bibinfo {author} {\bibfnamefont {M.~T.}\ \bibnamefont
  {Sieger}}, \bibinfo {author} {\bibfnamefont {T.}~\bibnamefont {Miller}}, \
  and\ \bibinfo {author} {\bibfnamefont {T.~C.}\ \bibnamefont {Chiang}},\
  }\href {\doibase 10.1103/PhysRevB.52.8256} {\bibfield  {journal} {\bibinfo
  {journal} {Physical Review B}\ }\textbf {\bibinfo {volume} {52}},\ \bibinfo
  {pages} {8256} (\bibinfo {year} {1995})}\BibitemShut {NoStop}%
\end{thebibliography}%

\end{document}